\documentclass[aps,prb,twocolumn,superscriptaddress]{revtex4}
\usepackage{graphicx}
\usepackage{subfigure}
\usepackage{amssymb,amsmath}
\usepackage{float}

\begin{document}

\bibliographystyle{apsrev}
\renewcommand{\floatpagefraction}{0.9}

\title{Molecular Dynamics Simulations of 
Evaporation-Induced Nanoparticle Assembly}

\author{Shengfeng Cheng}
\affiliation{Sandia National Laboratories, 
Albuquerque, NM 87185, USA}
\email{sncheng@sandia.gov}
\author{Gary S. Grest}
\affiliation{Sandia National Laboratories, 
Albuquerque, NM 87185, USA}

\date{\today}

\begin{abstract}
While evaporating solvent is a widely used technique to assemble
nano-sized objects into desired superstructures, 
there has been limited work on how the assembled structures
are affected by the physical aspects of the process.
We present large scale molecular dynamics simulations of
the evaporation-induced assembly of nanoparticles
suspended in a liquid that evaporates in a controlled fashion. 
The quality of the nanoparticle crystal formed just below the liquid/vapor interface 
is found to be better at 
relatively slower evaporation rates,
as less defects and grain boundaries appear.
This trend is understood as the result of the competition between the accumulation 
and diffusion times of nanoparticles at the liquid/vapor interface.
When the former is smaller,
nanoparticles are deposited so fast at the interface 
that they do not have sufficient time to arrange through diffusion,
which leads to the prevalence of defects and grain boundaries.
Our results have important implications in understanding assembly of
nanoparticles and colloids in non-equilibrium liquid environments.
\end{abstract}

\maketitle

\noindent {\bf I. INTRODUCTION}
\bigskip

Large, defect-free crystalline arrays of nanoparticles (NPs) are critical 
for many technologically important ultra thin film materials 
including filters, sensors, optical devices, and magnetic storage media.\cite{zhang10,he11} 
One common method for dispersing nanoparticles is to suspend them in solution, 
spread the suspension on a surface, 
and then evaporate the solvent.\cite{brinker99,bresme07,zhang10,pichumani12} 
This technique has been used to assemble NPs into
desired structures such as nanoclusters, rings, wires, stripes, films, and superlattices.
\cite{lin01,wyrwa02,liu02pccp,hoogenboom04,bigioni06,xu07,derkachov08,
chen09acsnano,xiao09,jeong10,he11,kanjanaboos11,dong11,fan12}
The factors that influence the assembled structure
include the evaporation kinetics of the solvent, the flow properties and concentration
of NPs in the solution, the NP-NP interactions,
the interactions between NPs and the liquid/vapor interface,
and the wetting/dewetting behavior of the suspension on the solid surface.\cite{jeong10}
One ubiquitous example is the "coffee-ring" stain left behind
after a particle-laden drop has evaporated.
A simple and beautiful physical picture of this class of phenomena was first
clarified by Deegan et al.\cite{deegan97} 
When the drop dries, a flow of liquid
from the interior of the drop to its edge emerges to replenish
the liquid that has evaporated from the region around the pinned contact line.
This flow convects particles to the drop's edge, 
where they form ring-like depositions.
The model was tested recently by Chen et al. on the molecular scale
with molecular dynamics (MD) simulations.\cite{koplik12}

Recently, Marin et al. demonstrated an order-to-disorder transition of 
the particle organization in coffee-ring stains.\cite{marin11,marin11POF}
In the early stage of evaporation, the deposition speed of particles
at the contact line is low and they have time to arrange 
into an ordered superlattice by Brownian diffusion.
However, at the end of evaporation, 
because of the increase of the flux velocity toward the contact line,
particles are deposited at a high speed and jammed into a quenched 
disordered phase as the result of ``rush-hour" traffic.
The experiment clearly showed the importance
of deposition/accumulation and diffusive time scales on controlling the particle ordering
during solvent evaporation.

Bigioni et al. showed earlier that particle accumulation and diffusion times
determine the assembly behavior 
and growth kinetics of monolayers of
gold NPs with diameters $\sim 6$ nm 
at a liquid/vapor interface.\cite{bigioni06,narayanan04,lin01} 
They found that in order to induce NP assembly at the interface,
the evaporation rate of the solvent has to be rapid 
and an attractive interaction between the NPs and interface is required.
The latter was induced by the excess of dodecanethiol in 
a toluene solution where gold NPs were suspended.
After the assembly is initiated, the growth kinetics of 
each crystalline domain can be exponential, sub-exponential,
or linear in time, depending on the ratio of a
diffusive length scale to the domain size.
By tuning the concentration of dodecanethiol, Bigioni et al. were able to make
long-range ordered arrays of gold NPs over macroscopic areas.\cite{bigioni06}

The work of of Bigioni et al. showed that evaporation generally
needs to be fast so that the accumulation of
NPs at the interface passes certain critical density, leading to the nucleation and
growth of 2-dimensional (2D) crystals. However,
if evaporation is too fast, 
the crystalline quality may diminish because
NPs do not have sufficient time to arrange via diffusion, but
instead quenched disordered structures will emerge,
similar to the ``rush-hour" effect observed in the coffee-ring stains.
This trend was actually observed earlier in an experiment 
of Im and Park using larger polystyrene colloids with diameters $\sim 230$ nm,\cite{im02} 
where the evaporation rate was tuned
by varying the temperature. They found
that there exists an optimum evaporation temperature 
that yields particle arrays of the highest quality at the water/air interface. 
When the evaporation temperature is higher than the optimum one,
the accumulation rate of colloids at the interface is too large,
leading to defects in the array.
It is thus interesting to determine if the effect
of evaporation rate on particle assembly persists to smaller NPs in the
fast-evaporation regime, which nicely fits the 
accessible length and time scales of MD simulations.
This was the motivation of our work reported here.
To anticipate the final results, 
we will show that there exits an optimum rate, 
determined by the diffusive times of NPs along the interface,
at which the 2D crystal of NPs formed near the interface has the highest quality
even without the ``annealing" step.

The remainder of this paper is organized as follows. The simulation methods 
are briefly introduced in Sec.~II. Then Sec.~III is devoted to results
and discussion. Conclusions are included in Sec.~IV.

\bigskip\noindent {\bf II. SIMULATION METHODS}
\bigskip

We modeled the liquid solvent and its vapor
with atoms interacting through a standard Lennard-Jones (LJ) potential,
$U_{\rm LJ}(r)=4\epsilon\left[ \left(\sigma/r\right)^{12}-\left(\sigma/r\right)^6\right]$,
where $r$ is the distance between two atoms,
$\epsilon$ the unit of energy, and $\sigma$ the diameter of atoms.
The potential was truncated and shifted to 0 at $r_c=3.0\sigma$.
The mass of each LJ atom is $m$ and a time unit
$\tau$ can be defined as $\tau=\sqrt{m\sigma^2/\epsilon}$.
The NPs of diameter $d=20\sigma$ were assumed to consist of a uniform
distribution of atoms interacting with a LJ potential.
For spherical particles, their mutual interaction
can be determined analytically
by integrating over all the interacting LJ atom pairs
between the two particles.\cite{Everaers03}
The resulting potential depends on $d$,
the inter-NP distance, and a Hamaker constant $A_{\rm nn}$
that characterizes the interaction strength.
We used a standard value $A_{\rm nn}=39.48\epsilon$ and 
a cut-off $20.427\sigma$.
This led to a purely repulsive NP-NP interaction,
which physically corresponds to adding a short surfactant
coating on NPs to avoid flocculation.\cite{veld09,grest11}
The interaction between LJ atoms and
NPs was determined similarly by integrating the interaction
between a single LJ atom and those within a NP.
We set the corresponding Hamaker constant 
$A_{\rm ns}=100\epsilon$, 
which with a cut-off of $14\sigma$ resulted in
NPs that were fully solvated and dispersed in the LJ solvent.\cite{cheng12JCP} 

\begin{figure*}[htb]
\centering
\includegraphics[width=6in]{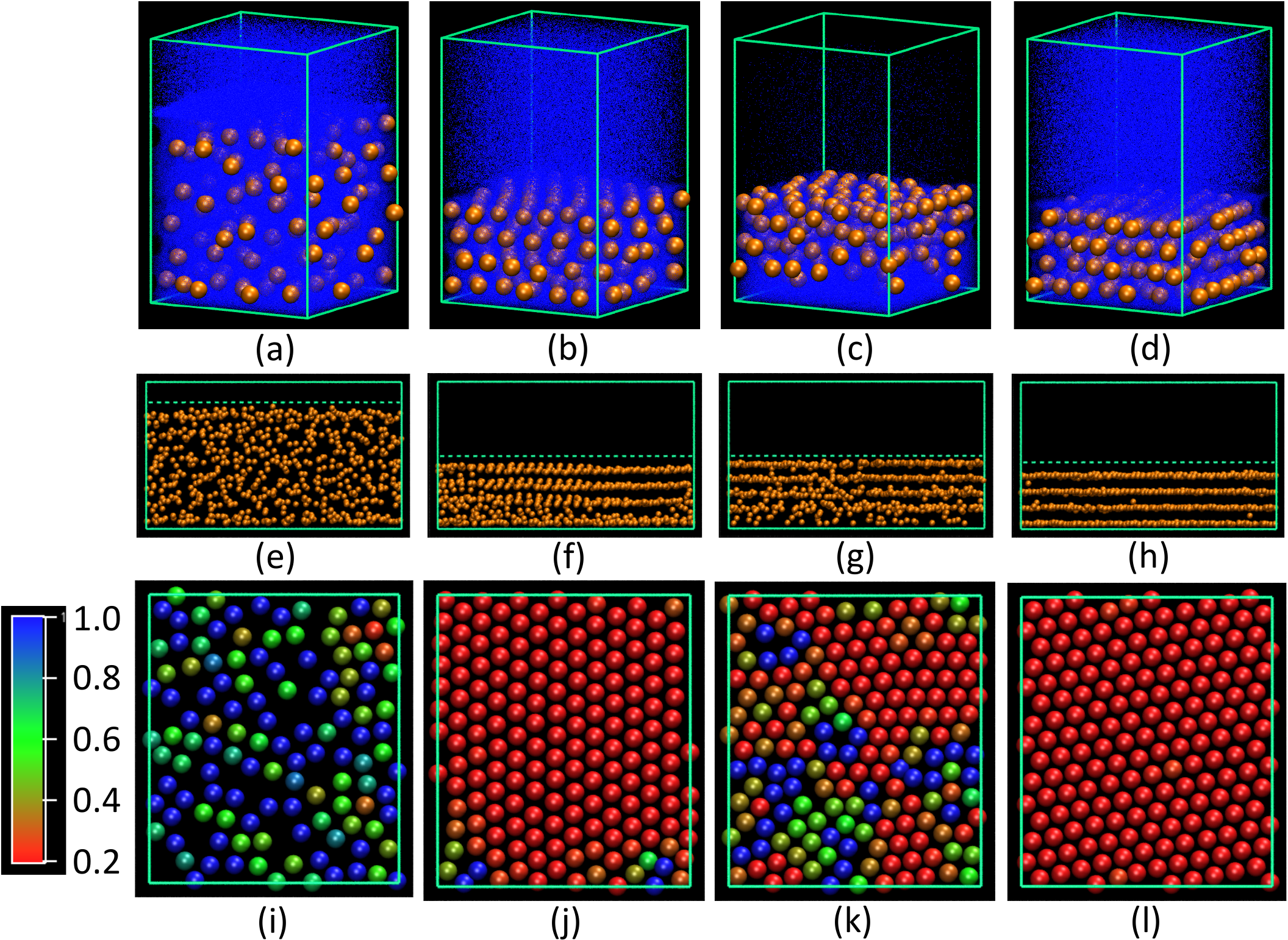}
\caption{Snapshots of a NP solution with $L_x=\sqrt{3}L_y/2=271\sigma$: 
(a) the equilibrated solution before the evaporation of solvent 
(containing $\sim 7.1$ million LJ atoms);
(b) after 52\% ($\sim 3.7$ million atoms) 
of the solvent had evaporated at a fixed rate 
$2.35\times 10^{-4}\tau^{-1}\sigma^{-2}$ 
for $1.85\times 10^5\tau$;
(c) after 52\% ($\sim 3.7$ million atoms)
of the solvent had evaporated into a vacuum 
for $2.9\times 10^4\tau$;
(d) the same system as in (c), 
but the system had re-equilibrated for $3.4\times 10^4\tau$ after
the evaporation was stopped.
For clarity in (a)-(d) only about a quarter of the system in the $x$-$y$ plane is shown.
(e)-(h): Side views of the NP distribution for systems in (a)-(d), respectively;
the dashed lines indicate the location of the liquid/vapor interface.
(i)-(l): Top views of NPs in the top layer for systems in (a)-(d), respectively; 
colors are based on an order parameter ($\overline{\psi_6}$) for each
NP as defined in Sec.~III(A).}
\label{assembly}
\end{figure*}

The simulation cell was a rectangular box of dimensions $L_x \times L_y \times L_z$.
The liquid/vapor interface was parallel to the $x$-$y$ plane,
in which periodic boundary conditions were employed.
In the $z$ direction, the LJ atoms and NPs were confined 
by two flat walls at $z=0$ and $z=L_z$, respectively.
The depth of the liquid was at least $150\sigma$
and the vapor was at least $70\sigma$ thick before evaporation.
The resulting liquid and vapor densities are $0.64m/\sigma^{3}$ and
$0.056m/\sigma^{3}$, respectively.
Note that LJ monomers have a vapor density higher than
most real fluids (Fig.~\ref{assembly}(a)).
This implies that the vaporization rate 
of LJ liquids is quite large and they can evaporate very fast.\cite{cheng11}
To provide a rough mapping of the evaporation rates 
(number fluxes of solvent atoms)
in our simulations to experimental values, we take
$\sigma \sim 0.3$ nm and $\tau \sim 2$ ps, 
the slowest evaporation rate we can simulate with MD is at the order
of $10^{-4}\tau^{-1}\sigma^{-2}$, corresponding to $\sim 5\times 10^{26} ~{\rm sec}^{-1}{\rm m}^{-2}$.
This is about an order of magnitude higher than the maximum rate at which
water evaporates, which is not surprising since water has
an unusually high boiling temperature and surface tension
compared with the LJ liquids studied here.

All simulations were performed using LAMMPS.\cite{plimpton95,lammps}
The equations of motion were integrated using a velocity-Verlet algorithm 
with a time step $\delta t =0.005\tau$.
During the equilibration, the temperature 
$T$ was held at $1.0\epsilon/k_{\rm B}$ by a Langevin thermostat weakly 
coupled to all LJ atoms with a damping constant $\Gamma=0.1\tau^{-1}$. 
Once the liquid/vapor interface was equilibrated, 
the Langevin thermostat was removed except
for those liquid atoms within $15\sigma$ 
of the lower wall at $z=0$.
We confirmed that the evaporation occurring at the interface
was not affected by the thermostat.\cite{cheng11}

\bigskip\noindent {\bf III. RESULTS AND DISCUSSION}
\bigskip

\noindent {\bf A. Effect of Evaporation Rate}
\bigskip

\begin{figure*}[htb]
  \subfigure[]{
    \begin{minipage}[b]{0.225\textwidth}
      \centering
      \includegraphics[width=1.55in]{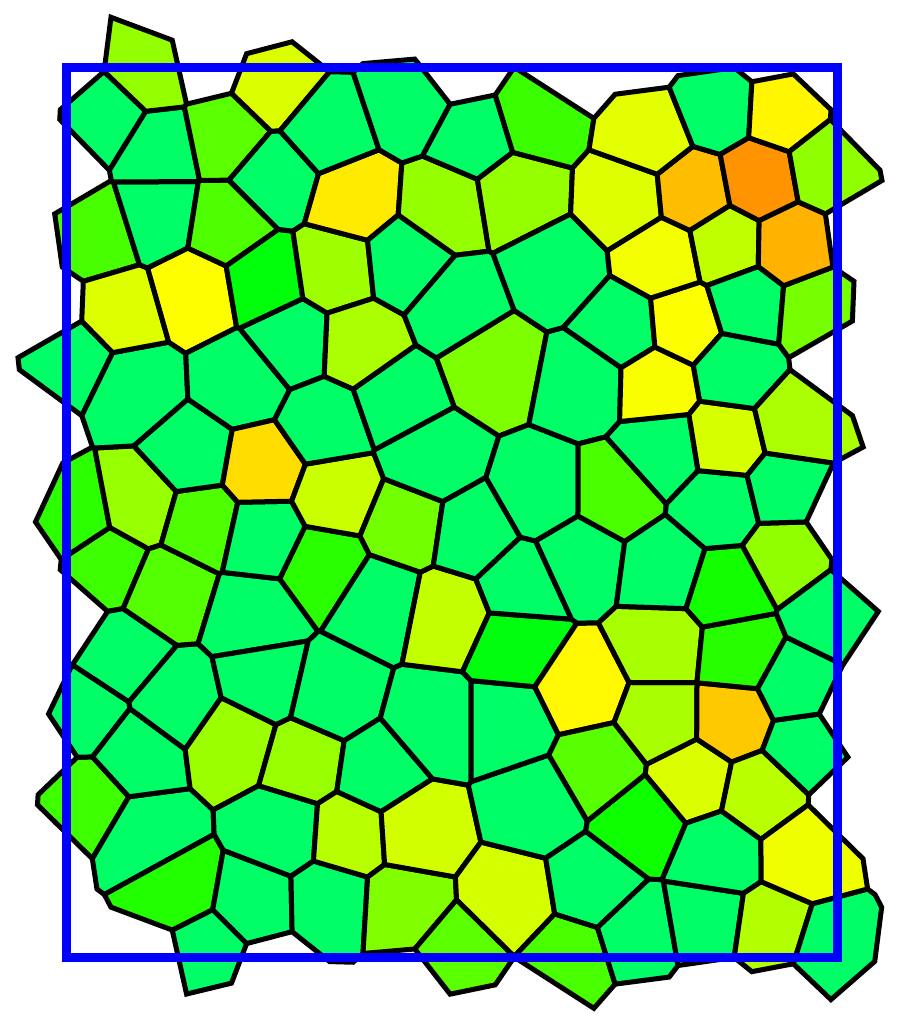}
     \end{minipage}}%
  \subfigure[]{
    \begin{minipage}[b]{0.225\textwidth}
      \centering
      \includegraphics[width=1.55in]{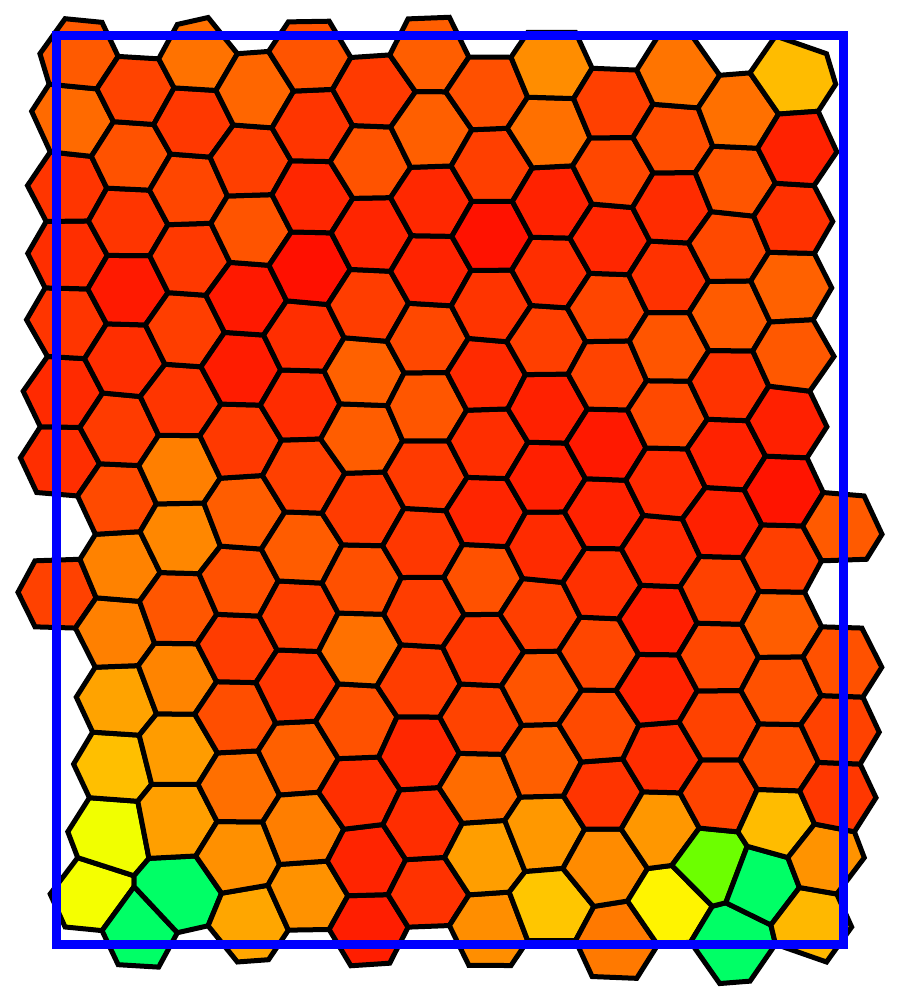}
     \end{minipage}}%
  \subfigure[]{
    \begin{minipage}[b]{0.225\textwidth}
      \centering
      \includegraphics[width=1.55in]{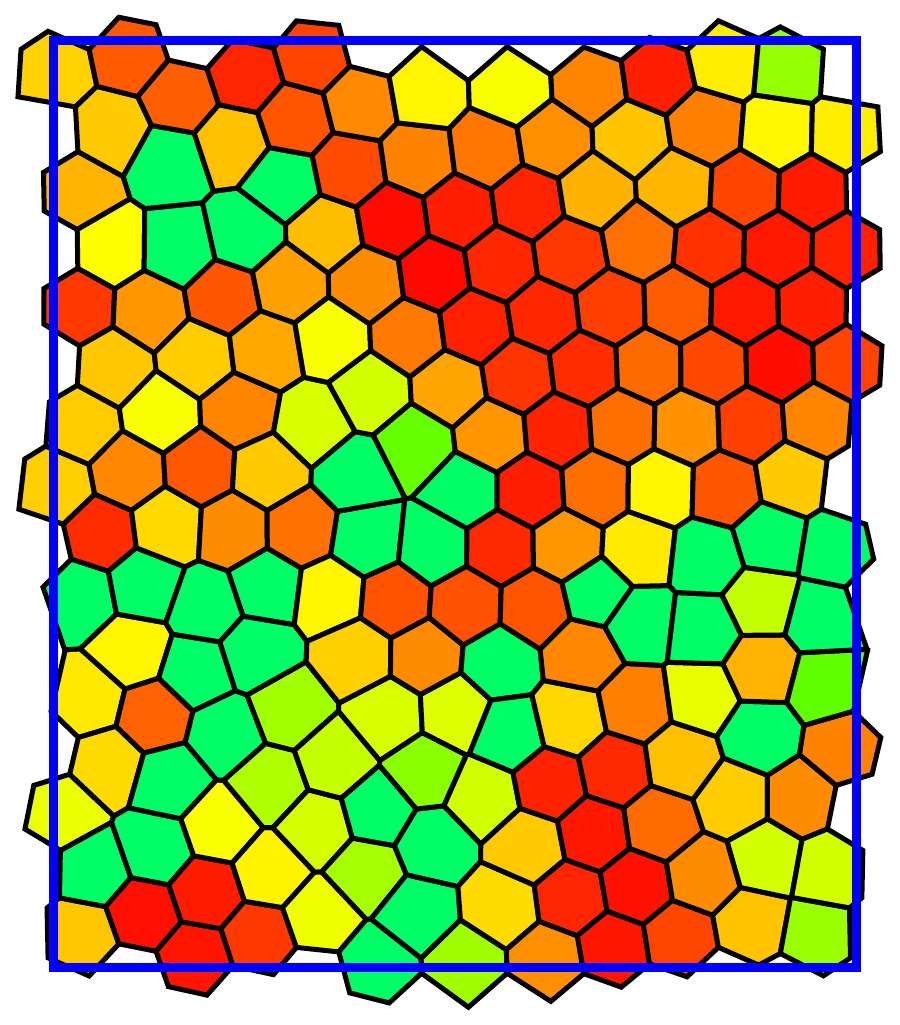}
     \end{minipage}} %
  \subfigure[]{
    \begin{minipage}[b]{0.225\textwidth}
      \centering
      \includegraphics[width=1.55in]{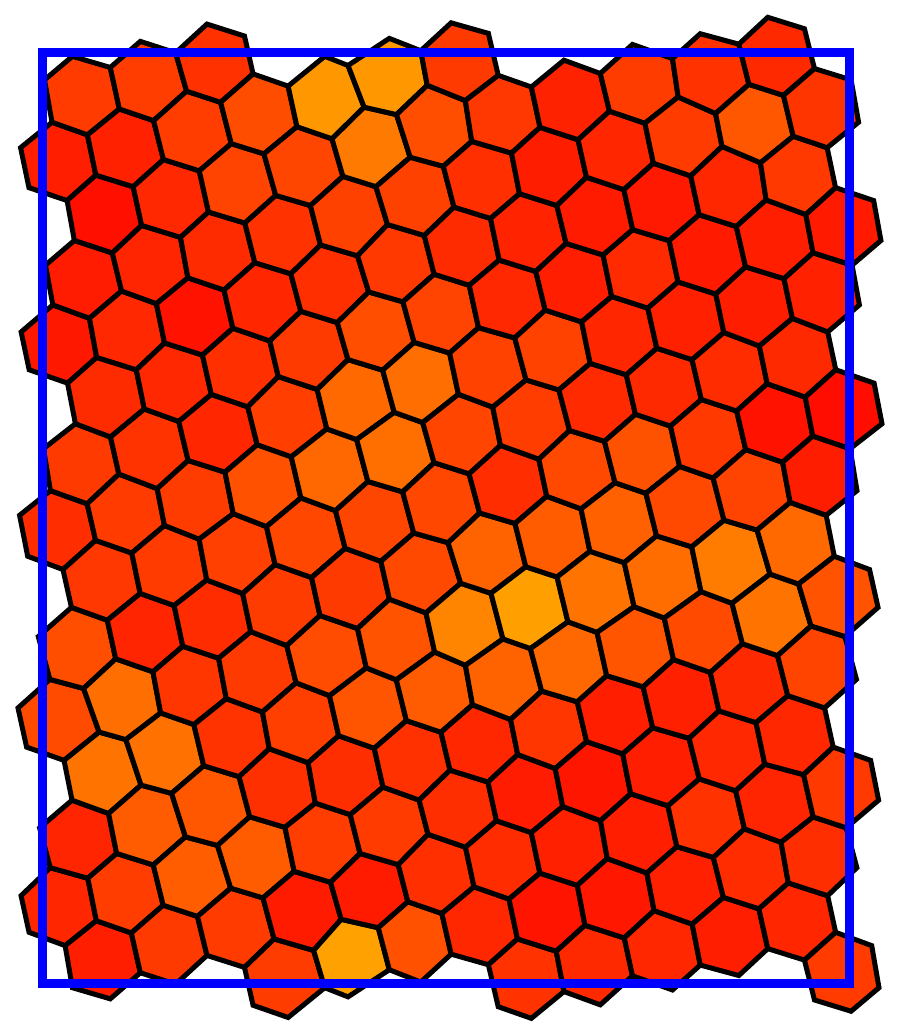}
     \end{minipage}}     
\caption{
(a)-(d): Voronoi constructions of NPs in the top layer in Figs.~\ref{assembly}(i), 
(j), (k), and (i), respectively.
Colors are based on $\overline{\psi_6}$.}
\label{voronoi_main}
\end{figure*}

Most of our simulations started with 668 NPs dispersed in a liquid
that contained about 7.1 million LJ atoms,\cite{note4} as shown in Fig.~\ref{assembly}(a).
The volume fraction of NPs was 0.205, which was based on the diameter of bare NPs. 
If we took into account the effect of a layer of solvent of thickness $\sim 1\sigma$
that was effectively attached to NPs,\cite{grest11} 
the effective volume fraction of NPs would be even higher,
roughly by a factor of 1.16.
We have studied systems of which the liquid/vapor interface has various aspect ratios.
Since the NPs were spherical and with a short-range
isotropic interaction, their ordering at the interface
was expected to be hexagonal.
Thus to reduce the finite size effects, 
we focused on a system with a cross-section that was commensurate with hexagonal 
packing, i.e., $L_x=\sqrt{3}L_y/2=271\sigma$ (Figs.~\ref{assembly} and \ref{voronoi_main}).
$L_x$ and $L_y$ were chosen so that the lattice spacing in a perfectly ordered hexagonal array
would be the same as that found after some of the solvent 
was evaporated in a system that had a square cross-section with 
$L_x=L_y=304\sigma$.
However, this latter system has a larger interfacial area than the previous one.
To reduce the potential influence of areal size, we
studied another system of which $L_y$ was increased to $352\sigma$ 
while $L_x$ was kept at $271\sigma$, 
so that not only the interfacial area was close to that of the square interface,
but also the aspect ratio of the interface favored a hexagonal packing 
with a lattice spacing found in the previous two systems.
Our results indicated that the size and aspect ratio of the interface
had negligible effects on the particle packing.
Below we mainly discuss the results for the first system
and include the results of other two systems in Supplementary Material.\cite{note5}

The side view of the NP distribution in the equilibrium solution 
is shown in Fig.~\ref{assembly}(e), where
the dashed line indicates the location of the liquid/vapor interface.
NPs are almost uniformly distributed in the solvent.
Note that during all our simulations, including the evaporation runs described below,
all NPs were fully solvated in the liquid solvent.
NPs within $d=20\sigma$ of
the interface were grouped together and referred as the top layer.
At equilibrium their areal coverage is about 45\%, which
is below the critical coverage 70\%
at the 2D hexatic phase transition.\cite{binder02}
As shown in Fig.~\ref{assembly}(i), the layer is clearly disordered.

To quantitatively characterize the in-plane packing geometry, we have defined
an order parameter $\psi_6$ for each NP.
A Voronoi construction was performed and those NPs with 6 neighbors were identified 
since in a close 2D packing each particle was expected to have 6 nearest neighbors
forming a hexagon. The angle 
$\theta_i$ corresponding to the $i$-th side of the hexagon was found
and $\psi_6$ was computed as
$\psi_6=\sum_{i=1}^{6}|\theta_i-\pi/6|$.
We found that $\psi_6$ was typically in the range of 
$0$ to $4\pi/3$. So we normalized $\psi_6$ by $4\pi/3$
and the normalized $\overline{\psi_6}$ was thus in the range of $0$ to $1$.
$\overline{\psi_6}=0$ indicates 
a perfect hexagonal packing,
while $\overline{\psi_6} \rightarrow 1$ indicates situations
far from a hexagonal lattice.
The Voronoi construction also showed that 
all other NPs had either 5 or 7 neighbors and they
were assigned $\overline{\psi_6}=1$ to indicate
that their packing configuration was not hexagonal.
In Fig.~\ref{assembly}(i)-(l), NPs are colored according
to their values of $\overline{\psi_6}$ 
(NPs with $\overline{\psi_6}\le 0.2$ are colored red and
with $\overline{\psi_6} = 1$ are colored blue). 
As shown in Fig.~\ref{assembly}(i), in an equilibrated solution
NPs near the interface are randomly distributed without any ordering,
which is also clear from the Voronoi construction shown in Fig.~\ref{voronoi_main}(a).

Evaporation of the solvent was modeled by introducing 
a deletion zone from $(L_z-20\sigma)$ to $L_z$. 
All or a certain number of vapor atoms
in the deletion zone were removed every $0.5\tau$. The former
mimicked evaporation into a vacuum which proceeded at the maximum rate, 
while the latter simulated evaporation at a fixed rate $j_E$,
namely the rate of removal of solvent atoms out of the simulation box.
Results in Figs.~\ref{assembly}(b), (f), and (j)
are for a system of which the solvent had evaporated 
at $j_E = 2.35\times 10^{-4}\tau^{-1}\sigma^{-2}$
since $t=0$.
During evaporation, the liquid film 
where NPs were dispersed shrunk, 
and NPs started to accumulate below the liquid/vapor interface
that impinged from above.
In this case the vapor density was only slightly reduced
compared with its equilibrium value (Fig.~\ref{assembly}(b))
throughout the whole simulation and the interface always remained 
close to equilibrium.
As a result, the rate $j_E$ was roughly the same as the actual outward
flux of solvent atoms near the liquid/vapor interface or in the vapor.
The side view of the NP distribution in Fig.~\ref{assembly}(f) 
shows that NPs formed layers where their concentration had peaks.
At $t=1.85\times 10^5\tau$, 52\% of the solvent ($\sim 3.7$ million atoms)
had evaporated, and the volume fraction of NPs increased to 0.35.
At least 3 layers of NPs
were formed as shown in Fig.~\ref{assembly}(f), with 171 NPs in the top layer.
Figure \ref{assembly}(j) shows the in-layer structure of these NPs,
from which a hexagonal packing is clearly visible.
This is also evident from the corresponding Voronoi construction (Fig.~\ref{voronoi_main}(b)).

Note that the number of layers induced by evaporation 
depends on the volume fraction of NPs
and the interfacial area.
In many experiments, NP solutions are quite dilute and
only one layer of NPs forms,
but more layers were also experimentally observed,\cite{lin01}
as in our simulations.
The structures in the NP layers below the top one were essentially similar but with a diminishing
crystalline quality as the distance from the interface increased.
Two consecutive layers were roughly in-registry as in fcc or hcp crystals,
but we had too few layers to distinguish the two crystal structures.
For clarity, we focus on the in-plane structure in the top NP layer below.

\begin{figure}[htb]
\centering
\includegraphics[width=3.25in]{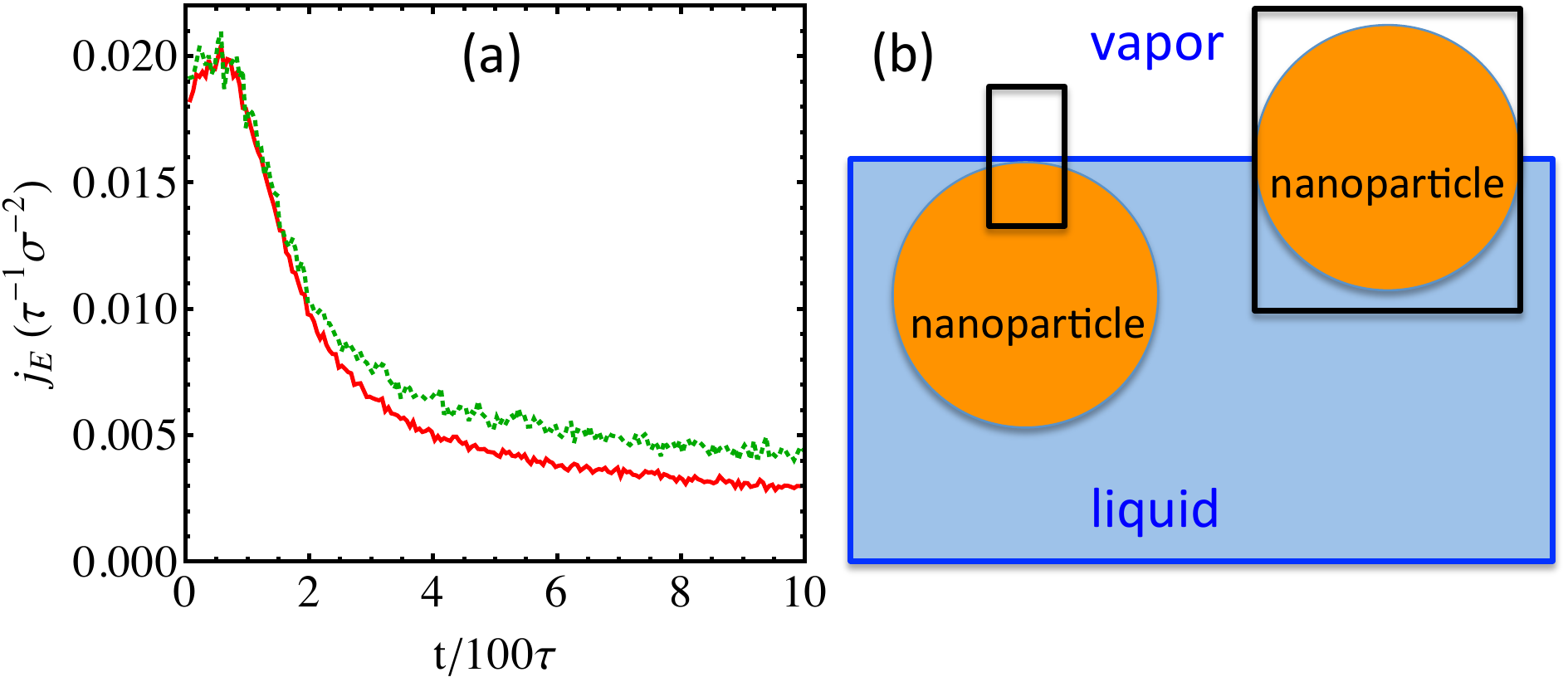}
\caption{(a) Evaporation rate $j_E$ vs. time for a NP solution (solid red line) 
and a neat solvent (dotted green line), 
showing the blockage effect of NPs on evaporation.
(b) A schematic illustration of how the evaporation
is affected by nanoparticles immersed just below or straddling the liquid/vapor interface:
only the evaporation of the solvent in the area enclosed by the black boxes is blocked.}
\label{evap_rate}
\end{figure}

The results in Figs.~\ref{assembly}(c), (g), and (k)
are for a system with the same starting state, but with the solvent evaporating into 
a vacuum since $t=0$.
Here the vapor was quickly depleted  (Fig.~\ref{assembly}(c)) and 
the evaporation at the liquid/vapor interface proceeded at the maximum rate,
which decreased with time and eventually reached a plateau value around 
$1.2\times 10^{-3}\tau^{-1}\sigma^{-2}$ (see Fig.\ref{evap_rate}). 
The actual flux of solvent atoms at the interface was lower than $j_E$
for the first $1000\tau$ during which the vapor depletion occurred, 
and then the two were almost identical.
It only took $2.9\times 10^4\tau$ to evaporate
roughly the same amount (52\%) of solvent as in the previous case with a fixed rate.
The layering of NPs was obvious from the side view (Fig.~\ref{assembly}(g)).
However, each layer was composed of several 2D grains separated
by grain boundaries, 
where the packing deviated significantly from a hexagonal lattice 
as indicated by colors based on the values of $\overline{\psi_6}$ 
(Fig.~\ref{assembly}(k)) and the corresponding Voronoi construction (Fig.~\ref{voronoi_main}(c))
for NPs in the top layer.

Comparing Figs.~\ref{assembly}(j) and (k) 
(or Figs.~\ref{voronoi_main}(b) and (c))
shows that the slower evaporation 
led to a NP packing with a better crystalline order.
The same trend was also found for the other two systems
evaporating at various rates (see Figs.~S1(a) and (b), 
and Figs.~S2(a) and (b) in Supplementary Material).\cite{note5}
This effect of evaporation rate $j_E$ on the assembly quality can
be understood on the basis of two time scales:
one describes how fast NPs accumulate at the interface, and
another is determined by the diffusivity of NPs along the interface.
In the steady state, the liquid/vapor interface recedes almost uniformly,
with a receding velocity $v$ that is related to
$j_E$ through $j_E=\rho v$, where $\rho=0.64\sigma^{-3}$ is the liquid density.
Using a model discussed by Bigioni {\it et. al.},\cite{bigioni06} a time scale
$t_a=(\phi v d^2)^{-1}$ can be defined to describe the speed of NP accumulation at the interface,
where $\phi$ is the NP concentration in the solution.
After NPs reach the top layer, they primarily diffuse 
along the interface. We denote the diffusive time scale as $t_d$, 
which can be estimated as $t_d=d^2/4D$ with $D$ as the diffusion coefficient.
For $d=20\sigma$ and $D \sim 3\times 10^{-3}\sigma^2/\tau$,\cite{cheng12JCP}
the diffusive time is $t_d \sim 3.3\times 10^4\tau$.

NPs accumulate faster than they diffuse in the limit $t_a \ll t_d$,
where the growth of 2D crystals is induced primarily by NP addition.
As it is unlikely that crystalline grains in different spatial regions have
orientations that match, this leads to the
occurrence of grain boundaries when grains meet.
However, in the opposite limit $t_a \gtrsim t_d$, 
NPs are added to the top layer slowly and small crystal regions can 
act as seeds which then grow larger through NP diffusion.
In this case defects and grain boundaries are less likely to appear.
The NP layer also has more time to relax into a uniform hexagonal lattice
even if imperfections in the packing occur.
For the system shown in Figs.~\ref{assembly}(k) and \ref{voronoi_main}(c), the solvent evaporated fast and 
the interface velocity $v$ was about $2\times 10^{-3}\sigma/\tau$,
which leads to $t_a \simeq 1.3 \sim 2.3 \times 10^4\tau$.
The value of $t_a$ has a range since
the NP concentration $\phi$ increased as the solvent evaporated.
For Figs.~\ref{assembly}(j) and \ref{voronoi_main}(b)
where $j_E$ was fixed at $2.35\times 10^{-4}\tau^{-1}\sigma^{-2}$, 
$v$ was about $3.7\times 10^{-4}\sigma/\tau$,
which gives $t_a \simeq 7 \sim 12 \times 10^4\tau$.
As expected, $t_a < t_d$ in the fast evaporation (into a vacuum), while 
$t_a > t_d$ in the slow evaporation (at a small fixed rate).

Since $t_a$ and $j_E$ are related through $t_a=(\phi d^2 j_E/\rho)^{-1}$,
by equating $t_a$ to $t_d$ one can estimate a critical rate $j_{C}=\rho/(\phi d^2 t_d)$.
When $j_E < j_{C}$, an almost defect-free hexagonal NP array is expected;
while for $j_E > j_{C}$, grain boundaries and point defects will 
appear in the 2D crystal induced by evaporation.
For our simulations, $j_{C} \simeq 9\times 10^{-4}\tau^{-1} \sigma^{-2}$.
When the solvent evaporated into a vacuum,
the evaporation rate was always larger than $j_{C}$,
leading to NP layers with defects and grain boundaries. 
However, when $j_E$ was reduced and fixed at a value below $j_{C}$,
the assembly quality in NP layers was greatly improved.
Furthermore, from experiments it is already known that 
if the evaporation is too slow, then the density of NPs at the interface
never reaches the critical nucleation density since NPs have abundant time
to diffuse back into the bulk of the solution, and assembly is not 
initiated in the first place.\cite{lin01, bigioni06}
Combined with this observation, our simulations indicate that there exits
an optimum evaporation rate $j_{\rm opt}$ at which the NP arrays
formed near the liquid/vapor interface are of the highest quality.
Though it is very difficult to precisely calculate $j_{\rm opt}$ from simulations,
our results indicate that $j_{\rm opt}$ should be about an oder of magnitude smaller
than $j_{C}$ since only a few defects and grain boundaries were observed
at $j_E\simeq j_{C}/5$ (Fig.~\ref{voronoi_main}(b) and Fig.~S1(a) in Supplementary Material),\cite{note5} 
while more were seen at $j_E\simeq j_{C}$ (Fig.~S2(a) in Supplementary Material).\cite{note5} 
Therefore, a reasonable estimate is $j_{\rm opt}\sim  j_{C}/10=\rho/(10\phi d^2 t_d)$.

\bigskip\noindent {\bf B. Effect of Annealing}
\bigskip

Since the volume fraction of NPs was only 0.35 in Fig.~\ref{assembly}(c) and
still below the critical fraction of hard-sphere crystallization ($\sim 0.545$),\cite{hoover68}
we expected that the layering structure in Fig.~\ref{assembly}(g) 
would be destroyed when the evaporation
was stopped and the system was allowed to relax. 
However, this was not the case on MD time scales, and the layering became even more
dramatic after the evaporation ceased at $t=2.9\times 10^4\tau$
and the system had re-requilibrated, 
as shown in Fig.~\ref{assembly}(d) and (h) at $t=6.3\times 10^4\tau$. 
The robustness of layering was partly due to the fact that some solvent evaporated
to fill the vapor-depleted region above the liquid and to re-establish
the liquid/vapor equilibrium. This increased the volume fraction of NPs in the solution to 0.38.
Furthermore, melting requires the diffusion of NPs in the direction perpendicular to the layers, 
which was hindered by the ordered structure in each layer and
the commensurability between layers.
Our results thus indicate that the system was dominated by non-equilibrium dynamics 
with long relaxation and equilibration times,
much larger than those achievable with MD simulations.

The organization of NPs in the top layer was greatly improved after relaxation. 
The grain boundaries in this layer completely disappeared after 
approximately $3.4\times 10^4\tau$
and NPs formed an almost perfect hexagonal lattice, 
as shown in Fig.~\ref{assembly}(l) and Fig.~\ref{voronoi_main}(d) 
(also see Figs.~S1(c) and S2(c) in Supplementary Material).\cite{note5}
This finding is consistent with an experimental observation that
alternative evaporation and relaxation produce
2D crystals of NPs with less defects.\cite{ray11}
It reveals the importance of relaxation that allows
imperfections in the NP packing to heal via structural re-arrangement,
which only gradually proceeds through the diffusion of NPs.
The improvement of the crystalline quality 
was also observed for the other layers below the top one after annealing, though
over the time scale of simulations they were still not completely ordered.
For NPs with diameters at the nanometer scale and 
without mutual attractions as studied here,
their diffusion is relatively fast and thus the time scale of structural
re-arrangement is accessible with MD.
However, it can be imagined that for larger NPs or colloidal particles
in the micrometer range and with attractive inter-particle interactions, 
diffusion becomes much slower
and it will take much longer time to remove defects and grain boundaries.
This is one of the reasons that they are frequently observed
in assembly experiments.\cite{lin01, bigioni06}

\bigskip\noindent {\bf C. Effect on Evaporation Rate of Assembled Nanoparticle Layers}
\bigskip

When NPs crystallize below the liquid/vapor interface,
they slow down the evaporation, since liquid atoms need
to circumvent NPs to reach the interface and then evaporate.
This blockage effect on evaporation is shown in Fig.~\ref{evap_rate}(a), 
where the rate $j_E$ is plotted against time for
a neat solvent and the NP solution that both evaporated into a vacuum.
At earlier times, NPs were dilute near the interface and liquid atoms
essentially did not feel their presence.
Therefore $j_E$ was roughly the same for the two systems.
At later times, since NPs started to form a dense layer below the interface, 
$j_E$ for the NP solution became smaller than that 
for the pure solvent. The factor of reduction was 25\%
when $j_E$ reached the plateau values in the two systems at $t\gtrsim 800\tau$.
However, the reduction in $j_E$ is not as dramatic as
what a naive estimate of the areal reduction of the liquid/vapor interface would indicate.
The reason is that NPs still sit below the interface in our simulations 
and their top sides only barely touched the liquid/vapor transition zone. 
As illustrated in Fig.~\ref{evap_rate}(b), 
only liquid atoms attached to the surface around the north pole of NPs were affected
in terms of evaporation. At several $\sigma$ away from this region 
(outside the box enclosing the north pole of the immersed NP in Fig.~\ref{evap_rate}(b)), the distance
from the interface to NP surface quickly increases beyond a few diameters of LJ atoms,
and the evaporation of the liquid there was not influenced by the presence of NPs.
However, if the NP layer straddled the liquid/vapor interface 
(see the straddling NP in Fig.~\ref{evap_rate}(b)),
we would expect $j_E$ to decrease by an order of magnitude compared with the case of a neat solvent.
In this case, the interfacial area is indeed reduced by NPs.
The evaporation cannot occur in the area covered by NPs 
and will slow down significantly.

\bigskip\noindent {\bf IV. CONCLUSIONS}
\bigskip

To summarize, we have studied the assembly of NPs in a solution induced by 
solvent evaporation with MD simulations. Our results showed that
NPs formed layers which started from below the liquid/vapor interface. 
The structure in each layer found in simulations is
similar to that observed in many experiments,\cite{lin01,narayanan04,bigioni06}
i.e., a close-packed hexagonal 2D crystal.
Interestingly, we found that the assembly quality was better when
the evaporation rate was relatively slower.
This trend was understood with a simple physical picture based on two time scales. 
The time scale associated with NP accumulation in the top layer, which
is controlled by the evaporation rate, needs to be larger than 
the diffusion time of NPs along the interface in order to form a defect-free 2D crystal. 
Otherwise, growth of crystals at different locations leads
to grains with different orientations, and grain boundaries and point defects
appear when grains meet.
Since it is known that evaporation needs to fast enough to initiate interfacial NP assembly in the first place, 
our simulations indicate that there exits
an optimum evaporation rate at which the packing quality of NP arrays
at the liquid/vapor interface is the best.
If evaporation proceeds at a higher rate, a relaxation process,
which can have a large time scale depending on the particle size and interaction,
is needed to ``anneal" the crystal and remove the imperfections in the packing.
Our simulations also showed the blockage effect on evaporation
because of the presence of the dense NP layer near the interface.
The effective area of the interface was reduced
and evaporation slowed down compared with a neat solvent.
However, in our simulations NPs were still immersed in the solvent and
only the solvent atoms in the region around
the top of NPs were actually affected in terms of evaporation and therefore only a minor reduction
in the evaporation rate was observed.

\section*{Acknowledgments}
This research used resources of the National Energy Research Scientific 
Computing Center (NERSC), which is supported by the Office of Science 
of the United States Department of Energy 
under Contract No. DE-AC02-05CH11231, and the Oak Ridge Leadership Computing Facility 
located in the National Center for Computational Sciences at Oak Ridge National Laboratory, 
which is supported by the Office of Science of the United States Department of Energy 
under Contract No. DE-AC05-00OR22725. 
These resources were obtained through the Advanced Scientific Computing Research (ASCR) 
Leadership Computing Challenge (ALCC).
This work is supported by the Laboratory Directed Research and 
Development program at Sandia National Laboratories.
Sandia National Laboratories is a multi-program laboratory 
managed and operated by Sandia Corporation, 
a wholly owned subsidiary of Lockheed Martin Corporation, 
for the U.S. Department of Energy's National Nuclear Security Administration 
under contract DE-AC04-94AL85000.

%\bibliography{allref}

\begin{thebibliography}{36}
\expandafter\ifx\csname natexlab\endcsname\relax\def\natexlab#1{#1}\fi
\expandafter\ifx\csname bibnamefont\endcsname\relax
  \def\bibnamefont#1{#1}\fi
\expandafter\ifx\csname bibfnamefont\endcsname\relax
  \def\bibfnamefont#1{#1}\fi
\expandafter\ifx\csname citenamefont\endcsname\relax
  \def\citenamefont#1{#1}\fi
\expandafter\ifx\csname url\endcsname\relax
  \def\url#1{\texttt{#1}}\fi
\expandafter\ifx\csname urlprefix\endcsname\relax\def\urlprefix{URL }\fi
\providecommand{\bibinfo}[2]{#2}
\providecommand{\eprint}[2][]{\url{#2}}

\bibitem[{\citenamefont{Zhang et~al.}(2010)\citenamefont{Zhang, Li, Zhang, and
  Yang}}]{zhang10}
\bibinfo{author}{\bibfnamefont{J.}~\bibnamefont{Zhang}},
  \bibinfo{author}{\bibfnamefont{Y.}~\bibnamefont{Li}},
  \bibinfo{author}{\bibfnamefont{X.}~\bibnamefont{Zhang}}, \bibnamefont{and}
  \bibinfo{author}{\bibfnamefont{B.}~\bibnamefont{Yang}},
  \bibinfo{journal}{Adv. Mater.} \textbf{\bibinfo{volume}{22}},
  \bibinfo{pages}{4249} (\bibinfo{year}{2010}).

\bibitem[{\citenamefont{He et~al.}(2011)\citenamefont{He, Lin, Chan,
  Vukovi\'{c}, Kr\'{a}l, and Jaeger}}]{he11}
\bibinfo{author}{\bibfnamefont{J.}~\bibnamefont{He}},
  \bibinfo{author}{\bibfnamefont{X.-M.} \bibnamefont{Lin}},
  \bibinfo{author}{\bibfnamefont{H.}~\bibnamefont{Chan}},
  \bibinfo{author}{\bibfnamefont{L.}~\bibnamefont{Vukovi\'{c}}},
  \bibinfo{author}{\bibnamefont{Kr\'{a}l}}, \bibnamefont{and}
  \bibinfo{author}{\bibfnamefont{H.~M.} \bibnamefont{Jaeger}},
  \bibinfo{journal}{Nano Lett.} \textbf{\bibinfo{volume}{11}},
  \bibinfo{pages}{2430} (\bibinfo{year}{2011}).

\bibitem[{\citenamefont{Brinker et~al.}(1999)\citenamefont{Brinker, Lu,
  Sellinger, and Fan}}]{brinker99}
\bibinfo{author}{\bibfnamefont{C.~J.} \bibnamefont{Brinker}},
  \bibinfo{author}{\bibfnamefont{Y.}~\bibnamefont{Lu}},
  \bibinfo{author}{\bibfnamefont{A.}~\bibnamefont{Sellinger}},
  \bibnamefont{and} \bibinfo{author}{\bibfnamefont{H.}~\bibnamefont{Fan}},
  \bibinfo{journal}{Adv. Mater.} \textbf{\bibinfo{volume}{11}},
  \bibinfo{pages}{579} (\bibinfo{year}{1999}).

\bibitem[{\citenamefont{Bresme and Oettel}(2007)}]{bresme07}
\bibinfo{author}{\bibfnamefont{F.}~\bibnamefont{Bresme}} \bibnamefont{and}
  \bibinfo{author}{\bibfnamefont{M.}~\bibnamefont{Oettel}},
  \bibinfo{journal}{J. Phys.: Condens. Matter} \textbf{\bibinfo{volume}{19}},
  \bibinfo{pages}{413101} (\bibinfo{year}{2007}).

\bibitem[{pic()}]{pichumani12}
\bibinfo{note}{{Pichumani M., Bagheri P., Poduska K. M., Vi\~{n}as W. G.,
  Yethiraj A., arXiv:1210.6662 [cond-mat.soft].}}

\bibitem[{\citenamefont{Lin et~al.}(2001)\citenamefont{Lin, Jaeger, Sorensen,
  and Klabund}}]{lin01}
\bibinfo{author}{\bibfnamefont{X.-M.} \bibnamefont{Lin}},
  \bibinfo{author}{\bibfnamefont{H.~M.} \bibnamefont{Jaeger}},
  \bibinfo{author}{\bibfnamefont{C.~M.} \bibnamefont{Sorensen}},
  \bibnamefont{and} \bibinfo{author}{\bibfnamefont{K.~J.}
  \bibnamefont{Klabund}}, \bibinfo{journal}{J. Phys. Chem. B}
  \textbf{\bibinfo{volume}{105}}, \bibinfo{pages}{3353} (\bibinfo{year}{2001}).

\bibitem[{\citenamefont{Wyrwa et~al.}(2002)\citenamefont{Wyrwa, Beyer, and
  Schmid}}]{wyrwa02}
\bibinfo{author}{\bibfnamefont{D.}~\bibnamefont{Wyrwa}},
  \bibinfo{author}{\bibfnamefont{N.}~\bibnamefont{Beyer}}, \bibnamefont{and}
  \bibinfo{author}{\bibfnamefont{G.}~\bibnamefont{Schmid}},
  \bibinfo{journal}{Nano lett.} \textbf{\bibinfo{volume}{2}},
  \bibinfo{pages}{419} (\bibinfo{year}{2002}).

\bibitem[{\citenamefont{Liu et~al.}(2002)\citenamefont{Liu, Zhu, Hu, and
  Liu}}]{liu02pccp}
\bibinfo{author}{\bibfnamefont{S.}~\bibnamefont{Liu}},
  \bibinfo{author}{\bibfnamefont{T.}~\bibnamefont{Zhu}},
  \bibinfo{author}{\bibfnamefont{R.}~\bibnamefont{Hu}}, \bibnamefont{and}
  \bibinfo{author}{\bibfnamefont{Z.}~\bibnamefont{Liu}},
  \bibinfo{journal}{Phys. Chem. Chem. Phys.} \textbf{\bibinfo{volume}{4}},
  \bibinfo{pages}{6059} (\bibinfo{year}{2002}).

\bibitem[{\citenamefont{Hoogenboom et~al.}(2004)\citenamefont{Hoogenboom,
  R\'{e}tif, de~Bres, van~de Boer, van Langen-Suurling, Romijn, and van
  Blaaderen}}]{hoogenboom04}
\bibinfo{author}{\bibfnamefont{J.~P.} \bibnamefont{Hoogenboom}},
  \bibinfo{author}{\bibfnamefont{C.}~\bibnamefont{R\'{e}tif}},
  \bibinfo{author}{\bibfnamefont{E.}~\bibnamefont{de~Bres}},
  \bibinfo{author}{\bibfnamefont{M.}~\bibnamefont{van~de Boer}},
  \bibinfo{author}{\bibfnamefont{A.~K.} \bibnamefont{van Langen-Suurling}},
  \bibinfo{author}{\bibfnamefont{J.}~\bibnamefont{Romijn}}, \bibnamefont{and}
  \bibinfo{author}{\bibfnamefont{A.}~\bibnamefont{van Blaaderen}},
  \bibinfo{journal}{Nano lett.} \textbf{\bibinfo{volume}{4}},
  \bibinfo{pages}{205} (\bibinfo{year}{2004}).

\bibitem[{\citenamefont{Bigioni et~al.}(2006)\citenamefont{Bigioni, Lin,
  Nguyen, Corwin, Witten, and Jaeger}}]{bigioni06}
\bibinfo{author}{\bibfnamefont{T.~P.} \bibnamefont{Bigioni}},
  \bibinfo{author}{\bibfnamefont{X.-M.} \bibnamefont{Lin}},
  \bibinfo{author}{\bibfnamefont{T.~T.} \bibnamefont{Nguyen}},
  \bibinfo{author}{\bibfnamefont{E.~I.} \bibnamefont{Corwin}},
  \bibinfo{author}{\bibfnamefont{T.~A.} \bibnamefont{Witten}},
  \bibnamefont{and} \bibinfo{author}{\bibfnamefont{H.~M.}
  \bibnamefont{Jaeger}}, \bibinfo{journal}{Nature Mater.}
  \textbf{\bibinfo{volume}{5}}, \bibinfo{pages}{265} (\bibinfo{year}{2006}).

\bibitem[{\citenamefont{Xu et~al.}(2007)\citenamefont{Xu, Xia, and Lin}}]{xu07}
\bibinfo{author}{\bibfnamefont{J.}~\bibnamefont{Xu}},
  \bibinfo{author}{\bibfnamefont{J.}~\bibnamefont{Xia}}, \bibnamefont{and}
  \bibinfo{author}{\bibfnamefont{Z.}~\bibnamefont{Lin}},
  \bibinfo{journal}{Angew. Chem. Int. Ed.} \textbf{\bibinfo{volume}{46}},
  \bibinfo{pages}{1860} (\bibinfo{year}{2007}).

\bibitem[{\citenamefont{Derkachov et~al.}(2008)\citenamefont{Derkachov, Kolwas,
  Jakubczyk, Zientara, and Kolwas}}]{derkachov08}
\bibinfo{author}{\bibfnamefont{G.}~\bibnamefont{Derkachov}},
  \bibinfo{author}{\bibfnamefont{K.}~\bibnamefont{Kolwas}},
  \bibinfo{author}{\bibfnamefont{D.}~\bibnamefont{Jakubczyk}},
  \bibinfo{author}{\bibfnamefont{M.}~\bibnamefont{Zientara}}, \bibnamefont{and}
  \bibinfo{author}{\bibfnamefont{M.}~\bibnamefont{Kolwas}},
  \bibinfo{journal}{J. Phys. Chem. C} \textbf{\bibinfo{volume}{112}},
  \bibinfo{pages}{16919} (\bibinfo{year}{2008}).

\bibitem[{\citenamefont{Chen et~al.}(2009)\citenamefont{Chen, Liao, Chen, Yang,
  Wark, Son, Batteas, and Cremer}}]{chen09acsnano}
\bibinfo{author}{\bibfnamefont{J.}~\bibnamefont{Chen}},
  \bibinfo{author}{\bibfnamefont{W.-S.} \bibnamefont{Liao}},
  \bibinfo{author}{\bibfnamefont{X.}~\bibnamefont{Chen}},
  \bibinfo{author}{\bibfnamefont{T.}~\bibnamefont{Yang}},
  \bibinfo{author}{\bibfnamefont{S.~E.} \bibnamefont{Wark}},
  \bibinfo{author}{\bibfnamefont{D.~H.} \bibnamefont{Son}},
  \bibinfo{author}{\bibfnamefont{J.~D.} \bibnamefont{Batteas}},
  \bibnamefont{and} \bibinfo{author}{\bibfnamefont{P.~S.}
  \bibnamefont{Cremer}}, \bibinfo{journal}{ACS Nano}
  \textbf{\bibinfo{volume}{3}}, \bibinfo{pages}{173} (\bibinfo{year}{2009}).

\bibitem[{\citenamefont{Xiao et~al.}(2009)\citenamefont{Xiao, Zhou, He, Li, and
  Yeung}}]{xiao09}
\bibinfo{author}{\bibfnamefont{L.}~\bibnamefont{Xiao}},
  \bibinfo{author}{\bibfnamefont{R.}~\bibnamefont{Zhou}},
  \bibinfo{author}{\bibfnamefont{Y.}~\bibnamefont{He}},
  \bibinfo{author}{\bibfnamefont{Y.}~\bibnamefont{Li}}, \bibnamefont{and}
  \bibinfo{author}{\bibfnamefont{E.~S.} \bibnamefont{Yeung}},
  \bibinfo{journal}{J. Phys. Chem. C} \textbf{\bibinfo{volume}{113}},
  \bibinfo{pages}{1209} (\bibinfo{year}{2009}).

\bibitem[{\citenamefont{Jeong et~al.}(2010)\citenamefont{Jeong, Hu, Lee,
  Garnett, Choi, and Cui}}]{jeong10}
\bibinfo{author}{\bibfnamefont{S.}~\bibnamefont{Jeong}},
  \bibinfo{author}{\bibfnamefont{L.}~\bibnamefont{Hu}},
  \bibinfo{author}{\bibfnamefont{H.~R.} \bibnamefont{Lee}},
  \bibinfo{author}{\bibfnamefont{E.}~\bibnamefont{Garnett}},
  \bibinfo{author}{\bibfnamefont{J.~W.} \bibnamefont{Choi}}, \bibnamefont{and}
  \bibinfo{author}{\bibfnamefont{Y.}~\bibnamefont{Cui}}, \bibinfo{journal}{Nano
  lett.} \textbf{\bibinfo{volume}{10}}, \bibinfo{pages}{2989}
  (\bibinfo{year}{2010}).

\bibitem[{\citenamefont{Kanjanaboos et~al.}(2011)\citenamefont{Kanjanaboos,
  Joshi-Imre, Lin, and Jaeger}}]{kanjanaboos11}
\bibinfo{author}{\bibfnamefont{P.}~\bibnamefont{Kanjanaboos}},
  \bibinfo{author}{\bibfnamefont{A.}~\bibnamefont{Joshi-Imre}},
  \bibinfo{author}{\bibfnamefont{X.-M.} \bibnamefont{Lin}}, \bibnamefont{and}
  \bibinfo{author}{\bibfnamefont{H.~M.} \bibnamefont{Jaeger}},
  \bibinfo{journal}{Nano Lett.} \textbf{\bibinfo{volume}{11}},
  \bibinfo{pages}{2567} (\bibinfo{year}{2011}).

\bibitem[{\citenamefont{Dong et~al.}(2011)\citenamefont{Dong, Ye, Chen, and
  Murray}}]{dong11}
\bibinfo{author}{\bibfnamefont{A.}~\bibnamefont{Dong}},
  \bibinfo{author}{\bibfnamefont{X.}~\bibnamefont{Ye}},
  \bibinfo{author}{\bibfnamefont{J.}~\bibnamefont{Chen}}, \bibnamefont{and}
  \bibinfo{author}{\bibfnamefont{C.~B.} \bibnamefont{Murray}},
  \bibinfo{journal}{Nano lett.} \textbf{\bibinfo{volume}{11}},
  \bibinfo{pages}{1804} (\bibinfo{year}{2011}).

\bibitem[{\citenamefont{Fan et~al.}(2012)\citenamefont{Fan, Bao, Sun, Bao,
  Manoharan, Nordlander, and Capasso}}]{fan12}
\bibinfo{author}{\bibfnamefont{J.~A.} \bibnamefont{Fan}},
  \bibinfo{author}{\bibfnamefont{K.}~\bibnamefont{Bao}},
  \bibinfo{author}{\bibfnamefont{L.}~\bibnamefont{Sun}},
  \bibinfo{author}{\bibfnamefont{J.}~\bibnamefont{Bao}},
  \bibinfo{author}{\bibfnamefont{V.~N.} \bibnamefont{Manoharan}},
  \bibinfo{author}{\bibfnamefont{P.}~\bibnamefont{Nordlander}},
  \bibnamefont{and} \bibinfo{author}{\bibfnamefont{F.}~\bibnamefont{Capasso}},
  \bibinfo{journal}{Nano lett.} \textbf{\bibinfo{volume}{12}},
  \bibinfo{pages}{5318} (\bibinfo{year}{2012}).

\bibitem[{\citenamefont{Deegan et~al.}(1997)\citenamefont{Deegan, Bakajin,
  Dupont, Huber, Nagel, and Witten}}]{deegan97}
\bibinfo{author}{\bibfnamefont{R.~D.} \bibnamefont{Deegan}},
  \bibinfo{author}{\bibfnamefont{O.}~\bibnamefont{Bakajin}},
  \bibinfo{author}{\bibfnamefont{T.~F.} \bibnamefont{Dupont}},
  \bibinfo{author}{\bibfnamefont{G.}~\bibnamefont{Huber}},
  \bibinfo{author}{\bibfnamefont{S.~R.} \bibnamefont{Nagel}}, \bibnamefont{and}
  \bibinfo{author}{\bibfnamefont{T.~A.} \bibnamefont{Witten}},
  \bibinfo{journal}{Nature} \textbf{\bibinfo{volume}{389}},
  \bibinfo{pages}{827} (\bibinfo{year}{1997}).

\bibitem[{kop()}]{koplik12}
\bibinfo{note}{{Chen W., Koplik J., Kretzschmar I., arXiv:1203.1910
  [cond-mat.soft].}}

\bibitem[{\citenamefont{\'{A}lvaro G.~Mar\'{i}n
  et~al.}(2011{\natexlab{a}})\citenamefont{\'{A}lvaro G.~Mar\'{i}n, Gelderblom,
  Lohse, and Snoeiger}}]{marin11}
\bibinfo{author}{\bibnamefont{\'{A}lvaro G.~Mar\'{i}n}},
  \bibinfo{author}{\bibfnamefont{H.}~\bibnamefont{Gelderblom}},
  \bibinfo{author}{\bibfnamefont{D.}~\bibnamefont{Lohse}}, \bibnamefont{and}
  \bibinfo{author}{\bibfnamefont{J.~H.} \bibnamefont{Snoeiger}},
  \bibinfo{journal}{Phys. Rev. Lett.} \textbf{\bibinfo{volume}{107}},
  \bibinfo{pages}{085502} (\bibinfo{year}{2011}{\natexlab{a}}).

\bibitem[{\citenamefont{\'{A}lvaro G.~Mar\'{i}n
  et~al.}(2011{\natexlab{b}})\citenamefont{\'{A}lvaro G.~Mar\'{i}n, Gelderblom,
  Lohse, and Snoeiger}}]{marin11POF}
\bibinfo{author}{\bibnamefont{\'{A}lvaro G.~Mar\'{i}n}},
  \bibinfo{author}{\bibfnamefont{H.}~\bibnamefont{Gelderblom}},
  \bibinfo{author}{\bibfnamefont{D.}~\bibnamefont{Lohse}}, \bibnamefont{and}
  \bibinfo{author}{\bibfnamefont{J.~H.} \bibnamefont{Snoeiger}},
  \bibinfo{journal}{Phys. Fluids} \textbf{\bibinfo{volume}{23}},
  \bibinfo{pages}{091111} (\bibinfo{year}{2011}{\natexlab{b}}).

\bibitem[{\citenamefont{Narayanan et~al.}(2004)\citenamefont{Narayanan, Wang,
  and Lin}}]{narayanan04}
\bibinfo{author}{\bibfnamefont{S.}~\bibnamefont{Narayanan}},
  \bibinfo{author}{\bibfnamefont{J.}~\bibnamefont{Wang}}, \bibnamefont{and}
  \bibinfo{author}{\bibfnamefont{X.-M.} \bibnamefont{Lin}},
  \bibinfo{journal}{Phys. Rev. Lett.} \textbf{\bibinfo{volume}{93}},
  \bibinfo{pages}{135503} (\bibinfo{year}{2004}).

\bibitem[{\citenamefont{Im and Park}(2002)}]{im02}
\bibinfo{author}{\bibfnamefont{S.~H.} \bibnamefont{Im}} \bibnamefont{and}
  \bibinfo{author}{\bibfnamefont{O.~O.} \bibnamefont{Park}},
  \bibinfo{journal}{Langmuir} \textbf{\bibinfo{volume}{18}},
  \bibinfo{pages}{9642} (\bibinfo{year}{2002}).

\bibitem[{\citenamefont{Everaers and Ejtehadi}(2003)}]{Everaers03}
\bibinfo{author}{\bibfnamefont{R.}~\bibnamefont{Everaers}} \bibnamefont{and}
  \bibinfo{author}{\bibfnamefont{M.~R.} \bibnamefont{Ejtehadi}},
  \bibinfo{journal}{Phys. Rev. E} \textbf{\bibinfo{volume}{67}},
  \bibinfo{pages}{041710} (\bibinfo{year}{2003}).

\bibitem[{\citenamefont{{in't Veld} et~al.}(2009)\citenamefont{{in't Veld},
  Petersen, and Grest}}]{veld09}
\bibinfo{author}{\bibfnamefont{P.~J.} \bibnamefont{{in't Veld}}},
  \bibinfo{author}{\bibfnamefont{M.~K.} \bibnamefont{Petersen}},
  \bibnamefont{and} \bibinfo{author}{\bibfnamefont{G.~S.} \bibnamefont{Grest}},
  \bibinfo{journal}{Phys. Rev. E} \textbf{\bibinfo{volume}{79}},
  \bibinfo{pages}{021401} (\bibinfo{year}{2009}).

\bibitem[{\citenamefont{Grest et~al.}(2011)\citenamefont{Grest, Wang, {in't
  Veld}, and Keffer}}]{grest11}
\bibinfo{author}{\bibfnamefont{G.~S.} \bibnamefont{Grest}},
  \bibinfo{author}{\bibfnamefont{Q.}~\bibnamefont{Wang}},
  \bibinfo{author}{\bibfnamefont{P.~J.} \bibnamefont{{in't Veld}}},
  \bibnamefont{and} \bibinfo{author}{\bibfnamefont{D.~J.}
  \bibnamefont{Keffer}}, \bibinfo{journal}{J. Chem. Phys.}
  \textbf{\bibinfo{volume}{134}}, \bibinfo{pages}{144902}
  (\bibinfo{year}{2011}).

\bibitem[{\citenamefont{Cheng and Grest}(2012)}]{cheng12JCP}
\bibinfo{author}{\bibfnamefont{S.}~\bibnamefont{Cheng}} \bibnamefont{and}
  \bibinfo{author}{\bibfnamefont{G.~S.} \bibnamefont{Grest}},
  \bibinfo{journal}{J. Chem. Phys.} \textbf{\bibinfo{volume}{136}},
  \bibinfo{pages}{214702} (\bibinfo{year}{2012}).

\bibitem[{\citenamefont{Cheng et~al.}(2011)\citenamefont{Cheng, Lechman,
  Plimpton, and Grest}}]{cheng11}
\bibinfo{author}{\bibfnamefont{S.}~\bibnamefont{Cheng}},
  \bibinfo{author}{\bibfnamefont{J.~B.} \bibnamefont{Lechman}},
  \bibinfo{author}{\bibfnamefont{S.~J.} \bibnamefont{Plimpton}},
  \bibnamefont{and} \bibinfo{author}{\bibfnamefont{G.~S.} \bibnamefont{Grest}},
  \bibinfo{journal}{J. Chem. Phys.} \textbf{\bibinfo{volume}{134}},
  \bibinfo{pages}{224704} (\bibinfo{year}{2011}).

\bibitem[{\citenamefont{Plimpton}(1995)}]{plimpton95}
\bibinfo{author}{\bibfnamefont{S.~J.} \bibnamefont{Plimpton}},
  \bibinfo{journal}{J. Comp. Phys.} \textbf{\bibinfo{volume}{117}},
  \bibinfo{pages}{1} (\bibinfo{year}{1995}).

\bibitem[{lam()}]{lammps}
\bibinfo{note}{{http://lammps.sandia.gov/}}.

\bibitem[{not({\natexlab{a}})}]{note4}
\bibinfo{note}{We also simulated a larger system starting with 720 NPs
  dispersed in a solvent containing about 17 million LJ atoms at a volume
  fraction 0.1. Results are qualitatively similar to those reported here, which
  indicates that our results are robust and insensitive to both the starting
  volume fraction of NPs and the finite size of the liquid/vapor interface.}

\bibitem[{not({\natexlab{b}})}]{note5}
\bibinfo{note}{See Supplementary Material Document No.XXXXX for Voronoi
  constructions of the other two systems with $L_x=L_y=304\sigma$, and
  $L_x=271\sigma$ and $L_y=352\sigma$, respectively. For information on
  Supplementary Material, see http://www.aip.org/pubservs/epaps.html.}

\bibitem[{\citenamefont{Binder et~al.}(2002)\citenamefont{Binder, Sengupta, and
  Nielaba}}]{binder02}
\bibinfo{author}{\bibfnamefont{K.}~\bibnamefont{Binder}},
  \bibinfo{author}{\bibfnamefont{S.}~\bibnamefont{Sengupta}}, \bibnamefont{and}
  \bibinfo{author}{\bibfnamefont{P.}~\bibnamefont{Nielaba}},
  \bibinfo{journal}{J. Phys.: Condens. Matter} \textbf{\bibinfo{volume}{14}},
  \bibinfo{pages}{2323} (\bibinfo{year}{2002}).

\bibitem[{\citenamefont{Hoover and Ree}(1968)}]{hoover68}
\bibinfo{author}{\bibfnamefont{W.~G.} \bibnamefont{Hoover}} \bibnamefont{and}
  \bibinfo{author}{\bibfnamefont{F.~H.} \bibnamefont{Ree}},
  \bibinfo{journal}{J. Chem. Phys.} \textbf{\bibinfo{volume}{49}},
  \bibinfo{pages}{3609} (\bibinfo{year}{1968}).

\bibitem[{ray()}]{ray11}
\bibinfo{note}{W. J. Ray, private communication.}

\end{thebibliography}

\section*{Supplementary Material}

In this Supplementary Material, we include Voronoi constructions of 
NPs in the top layer for the other 2 systems that we have studied.
They underwent various evaporation schemes,
as is detailed in the captions of each figure.
Figure~\ref{voronoi_np} is for a system with a square
cross section ($L_x=L_y=304\sigma$).
Figure~\ref{voronoi_solcon} is for another 
system that had a similar area, 
but the sizes of the liquid/vapor interface 
were adjusted to accommodate an integer number of NPs in both 
$x$ and $y$ directions ($L_x=271\sigma$ and $L_y=352\sigma$), respectively.

\setcounter{figure}{0}
\renewcommand{\thefigure}{S\arabic{figure}}
\begin{figure*}[htb]
   \subfigure[]{
   \begin{minipage}[b]{0.31\textwidth}
       \centering
       \includegraphics[width=2in]{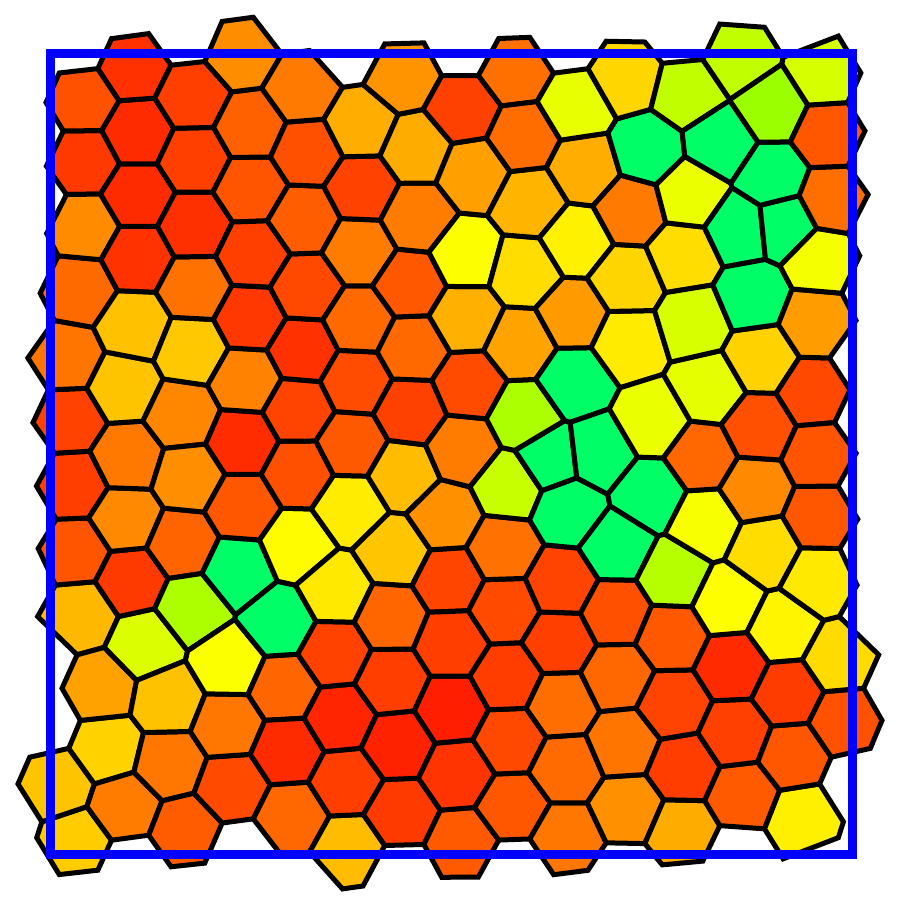}
   \end{minipage}}%
   \subfigure[]{
   \begin{minipage}[b]{0.31\textwidth}
       \centering
       \includegraphics[width=2in]{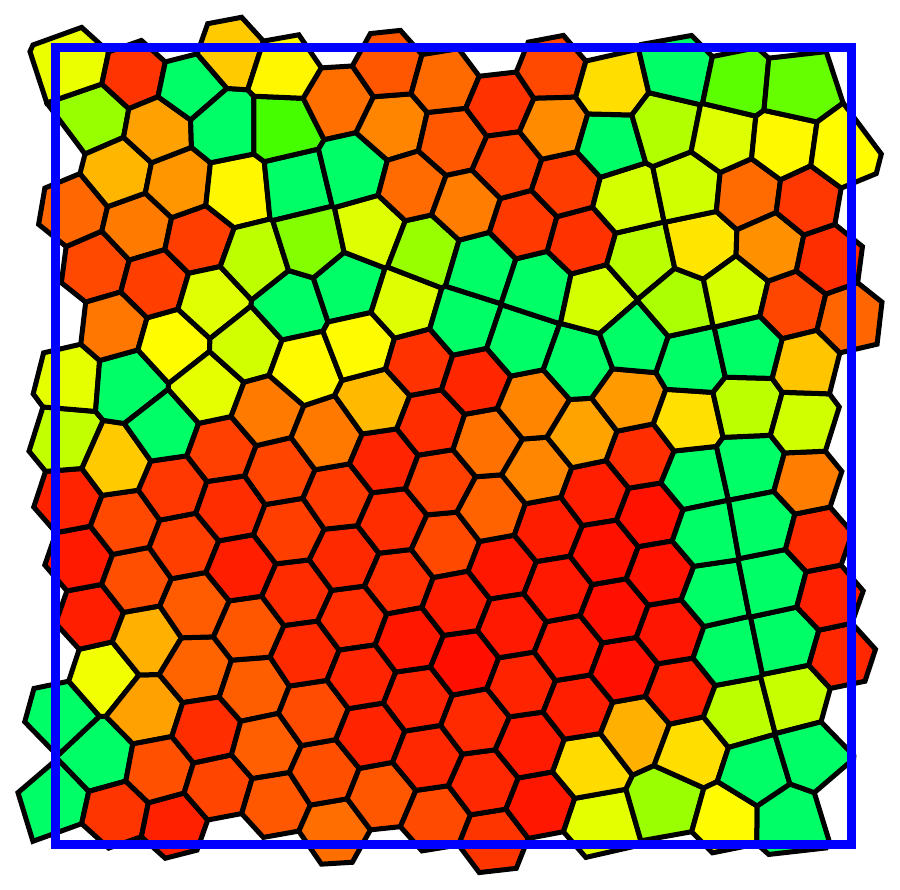}
   \end{minipage}}%
      \subfigure[]{
   \begin{minipage}[b]{0.31\textwidth}
       \centering
       \includegraphics[width=2in]{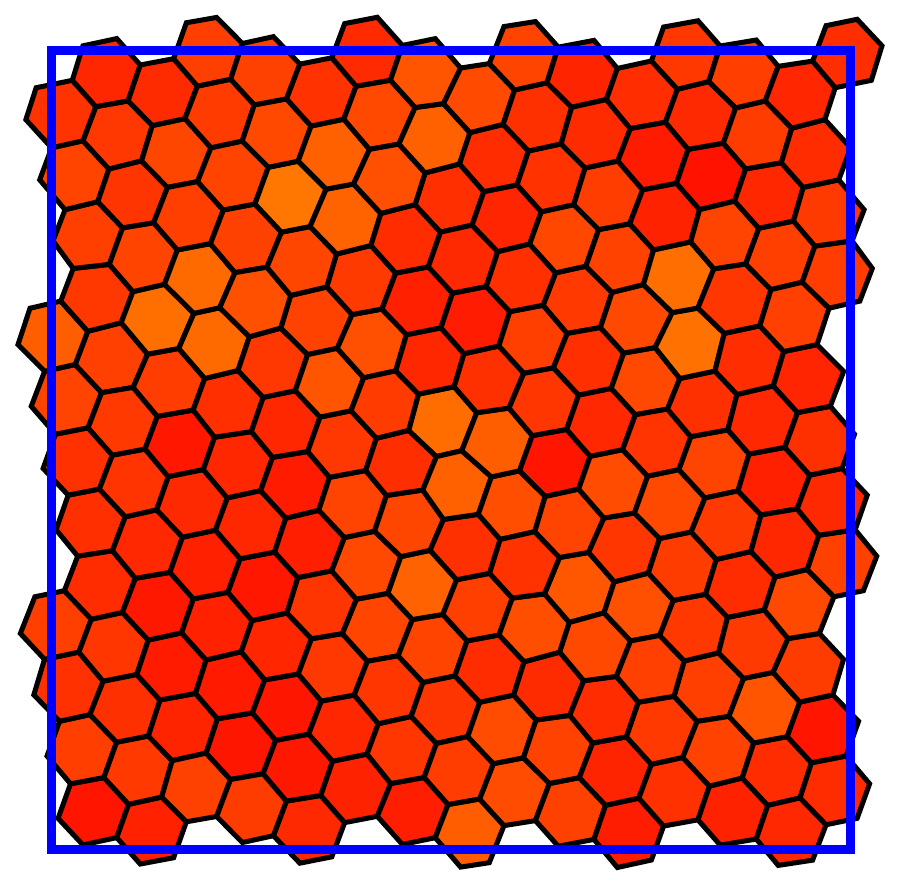}
   \end{minipage}}
\caption{Voronoi constructions of NPs in the top layer for the system
with $L_x=L_y=304\sigma$:
(a) after 45\% ($\sim 3.2$ million atoms) of the solvent had evaporated at 
a fixed rate $2.17\times 10^{-4}\tau^{-1}\sigma^{-2}$ 
for $1.645\times 10^5\tau$;
(b) after 45\% ($\sim 3.2$ million atoms) of the solvent 
had evaporated into a vacuum for $t=1.9\times 10^4\tau$;
(c) the same system as in (b) but relaxed for $3.4\times 10^4\tau$
after the evaporation was stopped.
}
\label{voronoi_np}
\end{figure*}

\begin{figure*}[htb]
    \subfigure[]{
    \begin{minipage}[b]{0.31\textwidth}
        \centering
        \includegraphics[width=2in]{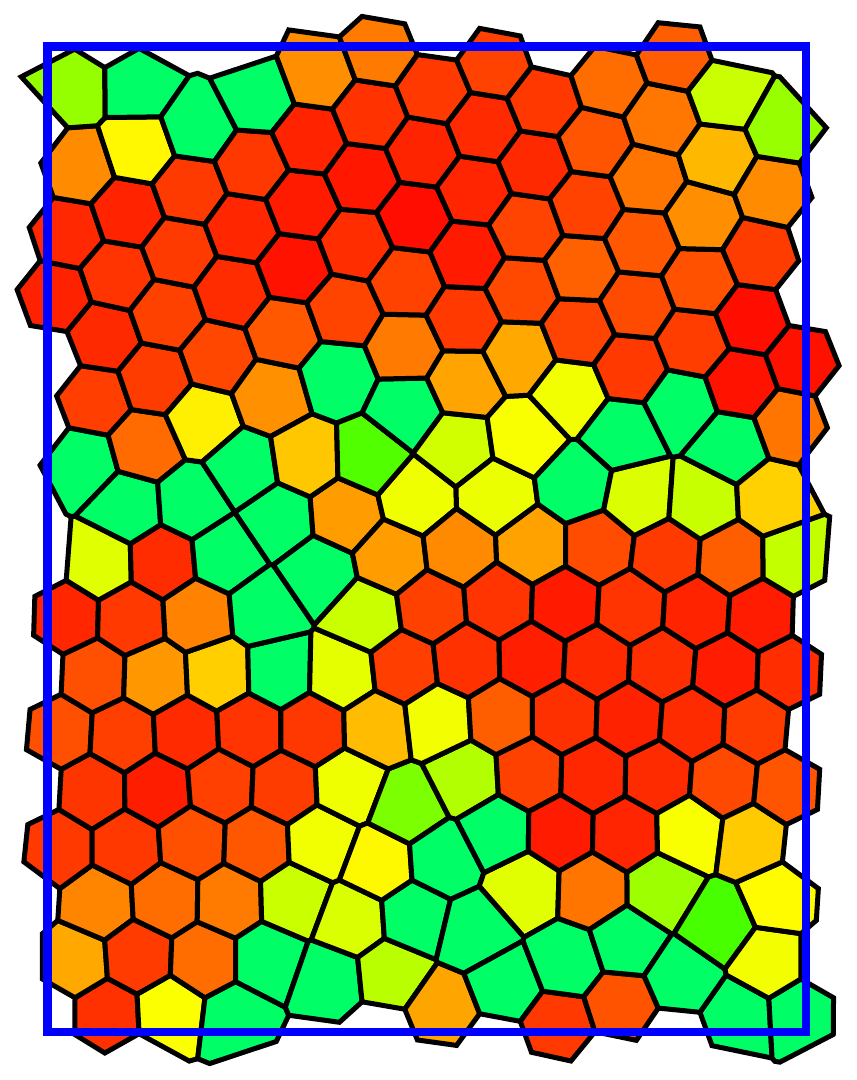}
    \end{minipage}}%
    \subfigure[]{
    \begin{minipage}[b]{0.31\textwidth}
        \centering
        \includegraphics[width=2in]{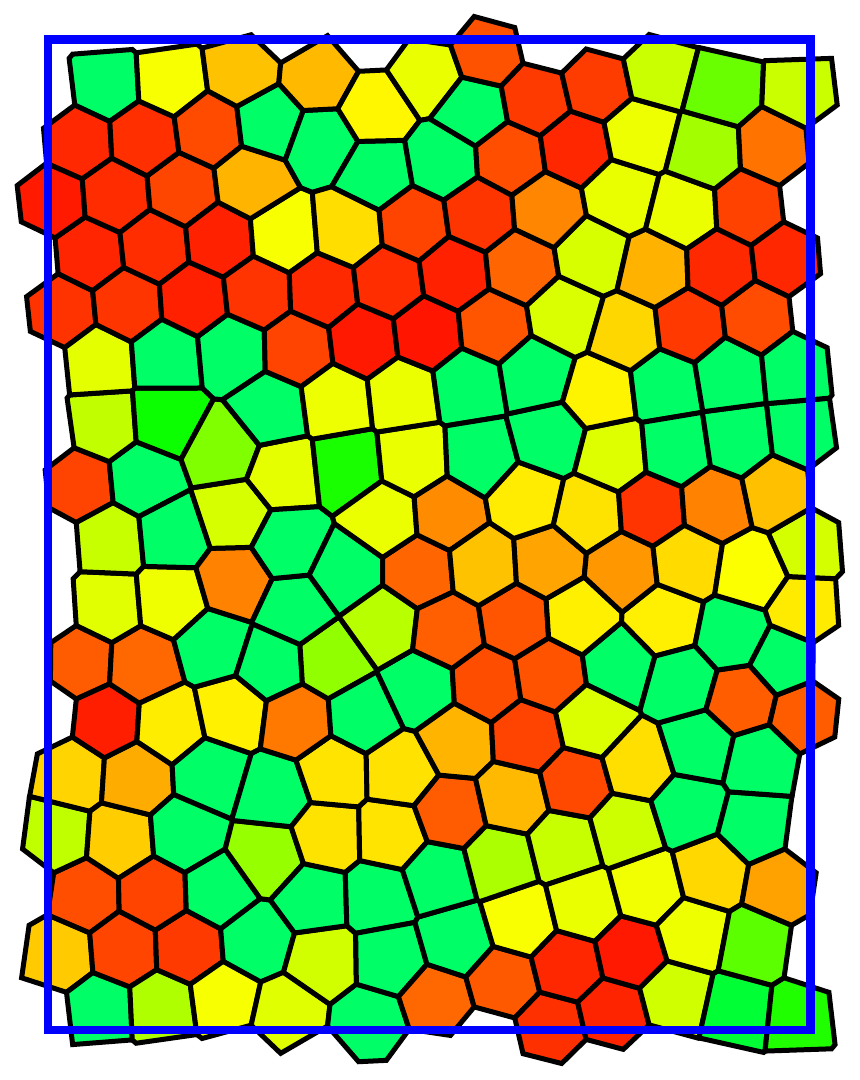}
    \end{minipage}}%
    \subfigure[]{
    \begin{minipage}[b]{0.31\textwidth}
        \centering
        \includegraphics[width=2in]{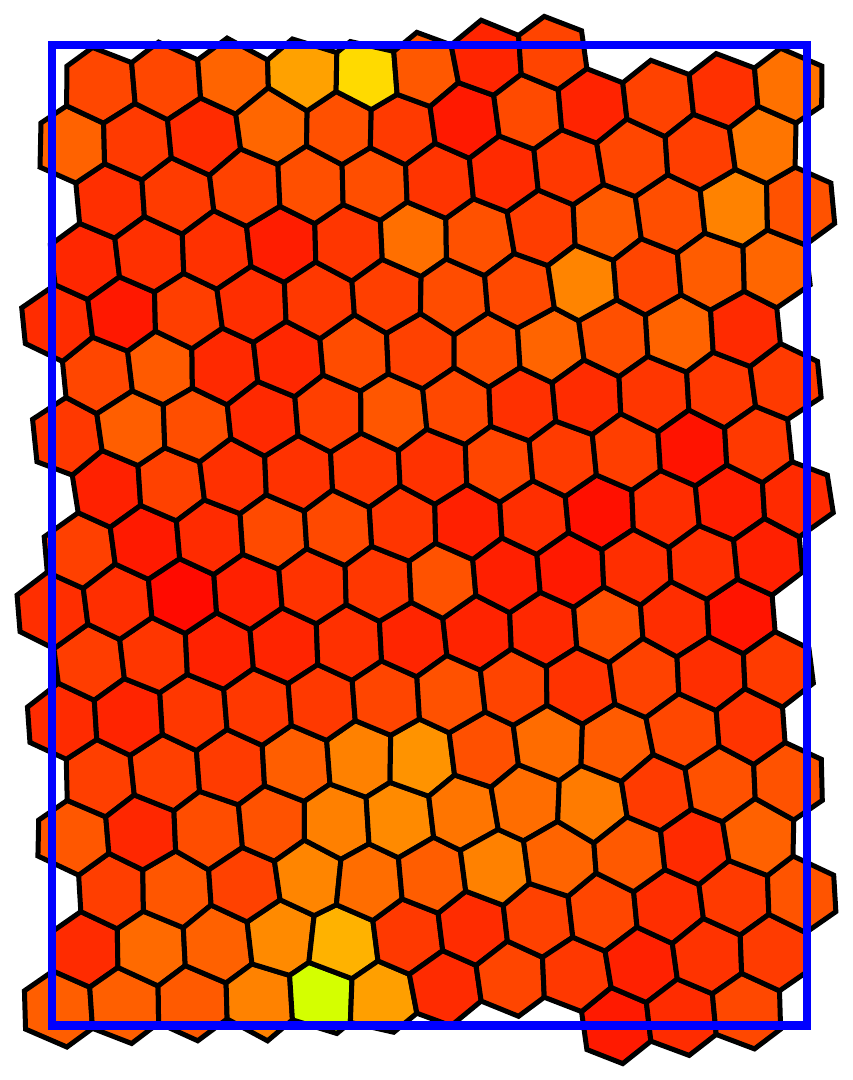}
    \end{minipage}}
\caption{Voronoi constructions of NPs in the top layer for the system with 
$L_x=271\sigma$ and $L_y=352\sigma$:
(a) after 42\% ($\sim 3$ million atoms) of the solvent 
had evaporated at a fixed rate $1.05\times 10^{-3}\tau^{-1}\sigma^{-2}$ 
for $3\times 10^4\tau$; 
(b) after 42\% ($\sim 3$ million atoms) of the solvent 
had evaporated into a vacuum for $1.6\times 10^4\tau$;
(c) the same system as in (b) but relaxed for $2.8\times 10^4\tau$
after the evaporation was stopped.
}
\label{voronoi_solcon}
\end{figure*}

\end{document}